\documentclass[aps,prc]{revtex4}

\newcommand{\be}{\begin{eqnarray}}
\newcommand{\ee}{\end{eqnarray}}

\newcommand{\q}{{\bf q}}
\newcommand{\p}{{\bf p}}

\def\bet{\mbox{\boldmath $\beta$}}
\def\pv{\mbox{\boldmath $p$}}
\def\p0v{\mbox{\boldmath $p_0$}}
\def\qv{\mbox{\boldmath $q$}}
\def\q0v{\mbox{\boldmath $q_0$}}

\usepackage{graphicx}
\usepackage{amsmath,amssymb,amsfonts}

\begin{document}


\title{Electromagnetic radiation from nuclear collisions at RHIC energies}

\author{Simon Turbide\footnote{Current address: Defence Research and Development Canada,  2459 Pie-XI Nord, Val-B\'elair, QC, Canada G3J 1X5}, Charles Gale}

\affiliation{Department of Physics, McGill University, 3600 University Street, Montr\'eal, QC, Canada H3A 2T8}

\author{Evan Frodermann$^1$, Ulrich Heinz$^{1,2}$}

\affiliation{$^{1}$Department of Physics, Ohio State University, Columbus, OH 43210, USA \\
$^{2}$CERN, Physics Department, Theory Division, CH-1211 Geneva 23, Switzerland}

\date{\today}

\begin{abstract}
The hot and dense strongly interacting matter created in collisions of heavy nuclei at RHIC energies is modeled with relativistic hydrodynamics, and the spectra of real and virtual photons produced at mid-rapidity in these events are calculated. Several different sources are considered, and their relative importance is compared. Specifically, we include jet fragmentation, jet-plasma interactions, the emission of radiation from the  thermal medium and from primordial hard collisions. Our calculations consistently take into account jet energy loss, as evaluated in the AMY formalism. We obtain results for the spectra, the nuclear modification factor ($R^{{\gamma}}_{AA}$), and the azimuthal anisotropy ($v_{2}^{\gamma}$) that agree with the photon  measurements performed by the PHENIX collaboration at RHIC. 
\end{abstract}

\maketitle

\section{Introduction}

As they interact only electromagnetically with the surrounding matter, real and virtual photons have the potential to probe the detailed dynamical history of high energy heavy ion collisions.   Their mean free path inside the hot and dense medium being much larger that its typical size, the photons will in principle leave the interacting zone without rescattering, reflecting directly the properties of the medium at the time they have been produced.  The photon is thus expected to be a good probe for the quark-gluon plasma (QGP), the search of which has driven many experiments over the last years. We concentrate in this paper on the conditions that prevail at the Relativistic Heavy Ion Collider (RHIC)~\cite{Phenix_white}, and see if the experimental results obtained there are amenable to a theoretical interpretation in terms of new physics.

Since the experimental detection of photons involves the entire collision, the QGP contribution might be hidden, or simply its effect reduced, by the sum of all other sources.  It it thus essential to have robust calculations for those contributions, which includes the photons produced during the overlap of the nuclei (prompt contribution), the hadron gas contribution, as well as the background coming from the decay of mesons ($\pi,\eta$), after the thermal freeze-out.  Considering that the background can experimentally be substracted, in principle by reconstructing the former mesons, we will concentrate on direct photons produced at RHIC in this paper.  Particularly, recent studies~\cite{prlphoton,simon2,Turbide:2006mc} have highlighted the role played by jets in real and virtual photon production.  In Ref.~\cite{TGF2006}, calculations have suggested that the direct interaction of jets with the QGP would generate an inverse anisotropy, which can be traduced in term of a negative coefficient $v_2$.  However, those results were obtained using a longitudinal expanding QGP.

In this work, the effect of the transverse expansion on photon production is evaluated. We use a 2D+1 hydrodynamical model, which has been applied recently with 
success to reproduce the characteristics of particle production at RHIC, 
such as momentum spectra, radial and elliptic flow. The spatial 
eccentricity in the model is also compatible with that inferred 
from experimental HBT measurements~\cite{Heinz:2005zg}. In Sec.\ref{sec_theory}, the various sources of photons are presented, and the way the transverse flow enter into their expressions is shown. The Sec.~\ref{sec_pro} presents the definition of all experimental observables that will be calculated, while in Sec.~\ref{sec_results}, the results are presented and analysed.  Finally, Sec.~\ref{sec_conclu} contains a summary and the conclusion.

\section{Photon production}
\label{sec_theory}

\subsection{Jet-thermal processes}

\begin{figure}
  \begin{center}
  \includegraphics[width=0.6\textwidth]{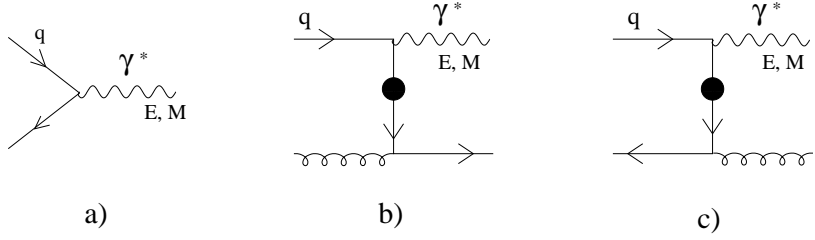}
  \end{center}
  \caption{\label{direct} Physical processes without collinear effects. }
\end{figure}

From finite-temperature field theory~\cite{vmd91}, the production rate of virtual photons with momentum $p$, invariant mass $M$ and energy $E$ is
\begin{equation}
\label{photon_TFT}
E\frac{dR^{\gamma^*}}{d^3p}(M)=\frac{1}{(2\pi)^3}\frac{\rm{Tr} \left[\mbox{Im}\,\Pi^{R}\right]}{1-e^{E/T}}\,,
\end{equation}
where {\rm Tr} $\left[ {\rm Im} \Pi^{R}\right] = {\rm Im} \Pi_\mu^{R\,\mu}$ is the trace of the imaginary part of the retarded photon self-energy. In the hard thermal loop (HTL) resummation formalism~\cite{htl}, the non-collinear processes contribution up to next to leading order in $g_s$ are shown in Fig.~\ref{direct}.  The filled circles in this figure indicate resummed propagators.   From relativistic kinetic theory, the production rate of those non-collinear processes induced by jets have the following form:
\begin{equation}
\label{ph_noncoll_rate}
E\frac{dR^{\gamma^*}_{\rm non-coll}}{d^3p}(M)= N_cN_s \int \frac{d^3q}{(2\pi)^3} f_{q+\bar{q}}^{\rm jet}(q,b) E\frac{d \Gamma^{q\to\gamma}_{\rm non-coll}(\qv,\pv,M)}{d^3p dt}\,.
\end{equation}
  The color and spin degeneracy factors are respectively $N_c=3$ and $N_s=2$.   The phase-space distribion of incoming jets, initially created at position ${\bf r}_\perp$, assuming a Bjorken $\eta-y$ correlation~\cite{lin} is
\be
f_{q+\bar{q}}^{\rm jet}({\bf x},\qv,t,b)=\frac{(2\pi)^3{\cal P}({\bf r}_\perp)}{6\tau q_{T}}\frac{dN_{q+\bar{q}}^{\rm jet}(t,b)}{d^2q_{T}dy}\delta(\eta - y)\,,
\ee
where $\tau=\sqrt{t^2-z^2}$ is the proper time, the $z$-axis being the beam direction, and $\eta=1/2 \mbox{ln}[(t+z)/(t-z)]$ is the space-time rapidity. 
The initial profile of jets in the transverse plane is obtained by
\be
{\cal P}({\bf r}_\perp,b)=\frac{T_A({\bf r}_\perp+\frac{{\bf b}}{2})T_A({\bf r}_\perp-\frac{{\bf b}}{2})}{T_{AB}(b)}\,,
\ee
where $T_A$ and $T_{AB}$ are the thickness and overlap functions, which are evaluated with a realistic Woods-Saxon distribution\cite{KH}.
  The initial momentum distribution of jets, at a given impact parameter $b$, is obtained by
\be
\label{jet_initial}
\left.\frac{dN_{\rm jet}(Q,b)}{d^2q_T dy}\right|_{t=0}&=&T_{AB}(b)\sum_{a,b,c}\int dx_a g_a(x_a,Q)g_b(x_b,Q)\nonumber \\ &&\times K \frac{d\sigma^{a+b\rightarrow c+jet}}{dt}\frac{2x_ax_b}{\pi
(2x_a-2\frac{q_T}{\sqrt{s}}e^{y})} 
\ee
where $\sqrt{s}=200$ GeV.  Isospin effects are included in the parton distribution function (pdf) by
\be
\label{iso_sha}
g_a(x_a,Q)=\left(\frac{Z}{A}f_a(x_a,Q)+\frac{A-Z}{A}f_{a^*}(x_a,Q)\right) R(x_a,Q)\,,
\ee
where $f_a$ is the parton distribution function inside proton~\cite{pdf}.  The second term, corresponding to pdf inside the neutron, is obtained by the following substitution of parton species: $a^*=(d,u,\bar{d},\bar{u},s,\bar{s},g)$ for $a=(u,d,\bar{u},\bar{d},s,\bar{s},g)$.  Shadowing effects are included in the function $R(x_a,Q)$ ~\cite{shadowing}. The factorization scale $Q$ is assumed to be $q_T$. We use a NLO factor $K=1.7$, which, according to Ref.~\cite{barnafoldi}, is almost $q_T$-independent. The jet distributions evolve in time according to~\cite{Jeon-Moore}
\be
\frac{dN^{\rm jet}_{q+\bar{q}}(q)}{dq_Tdydt} & = & \int_k \!
        \frac{dN^{\rm jet}_{q+\bar{q}}}{dq_T dy} (q{+}k) \frac{d\Gamma^q_{\!qg}(q{+}k,k)}{dkdt}
   -\frac{dN^{\rm jet}_{q+\bar{q}}}{dq_Tdy} (q)\frac{d\Gamma^q_{\!qg}(q,k)}{dkdt}
        \nonumber \\ &&
        +2\frac{dN^{\rm jet}_g}{dq_Tdy} (q{+}k)\frac{d\Gamma^g_{\!q \bar q}(q{+}k,k)}{dkdt} \, , \nonumber \\
\frac{dN^{\rm jet}_g (q)}{dq_Tdydt} &  = & \int_k \!
        \frac{dN^{\rm jet}_{q+\bar{q}}}{dq_T dy} (q{+}k) \frac{d\Gamma^q_{\!qg}(q{+}k,q)}{dkdt}
        {+}\frac{dN^{\rm jet}_g}{dq_Tdy} (q{+}k)\frac{d\Gamma^g_{\!\!gg}(q{+}k,k)}{dkdt}
        \nonumber \\ && 
        -\frac{dN^{\rm jet}_g}{dq_Tdy}(q) \left(\frac{d\Gamma^g_{\!q \bar q}(q,k)}{dkdt}
        \right.+\left.\frac{d\Gamma^g_{\!\!gg}(q,k)}{dkdt} \Theta(2k{-}q) \!\!\right) ,\nonumber \\
\label{eq:Fokker}
\ee
where the $k$ integrals run from $-\infty$ to $\infty$.   The transition rates in the laboratory frame are  
\be
\frac{d\Gamma^q_{qg}(q,k)}{dkdt}=(1-{\bf v_{jet}}\cdot\bet)\,\frac{d\Gamma^q_{qg}(q_{0},k_0)}{dk_0dt_0}\,,
\ee
where $d\Gamma^q_{qg}/dk_0dt_0$ are evaluated in the fluid local frame moving with a velocity $\mbox{\boldmath $\beta$}$ relatively to the laboratory frame, and $(1-{\bf v_{jet}}\cdot\bet)$ represents the Jacobian of the $dt_0\,dk_0\rightarrow dt\,dk$ transformation.  The transition rates in the local thermal frame are taken from the AMY~\cite{AMY} formalism, and include radiative energy loss through gluon bremsstrahlung with LPM effect.  Processes like $q \bar{q}$ annihilation and absorption of thermal gluons are also included in the model.  The strength of the transition rates is controlled by the strong coupling constant $\alpha_s$ and the temperature $T$. The temperature dependence of $\alpha_s$ is obtained from lattice QCD~\cite{lat_alph}. As the jets propagate in the QGP, the parameters $\bet$ and $T$ which depend on the position of the jets and the time, are directly extracted from the hydro model. All jet-medium interactions cease when the critical temperature $T_c$ is reached. The jets distribution, appearing in Eq.~(\ref{ph_noncoll_rate}), is evaluated in the laboratory frame, so must be the quark to photon transition rates.    Since $E_0/d^3p_0$ is a Lorentz invariant, the Jacobian for the Lorentz transformation is simply $\partial t_0/\partial t=\sqrt{1-|\bet|^2}$. We thus get
\be
E\frac{d \Gamma^{q\to\gamma}_{\rm non-coll}(\qv ,\pv)}{d^3p dt}=\sqrt{1-|\bet|^2}E_0\frac{d \Gamma^{q\to\gamma}_{\rm non-coll}(\qv_{0},\pv_0,M)}{d^3p_0 dt_0}\,.
\ee
The quark to photon transition rates in the local thermal frame have been calculated in Ref.~\cite{Turbide:2006mc}.  The photon's momentum and energy in that frame are
\be
\pv_0=\left[\frac{\pv\cdot\bet-|\bet|^2 E}{|\bet|\sqrt{1-|\bet|^2}}\right]\frac{\bet}{|\bet|}+\left[\pv-\pv\cdot\bet\frac{\bet}{|\bet|^2}\right]\,,\quad\quad E_0=\sqrt{p_0^2+M^2}\,.
\ee
The jet's momentum follows the same transformation rule, but with $E_q=q$ since we assume the jets to be massless $(|{\bf v_{jet}}=1|)$.
The yield of virtual photons produced in non-collinear processes induced by jets in the expanding medium is 
\be
E\frac{dN^{\gamma^*}_{\rm non-coll}}{d^3p}(M)&=&\int d^4x\,E\frac{dR^{\gamma^*}}{d^3p}(M)\nonumber \\ &=&\int d\tau\tau\, d\eta\, d^2x_\perp\,N_cN_s \int \frac{d^3q}{(2\pi)^3} f_{q+\bar{q}}^{\rm jet}(q,b) \sqrt{1-\beta(\tau,\eta,{\bf x}_\perp)^2}\,E_0\frac{d \Gamma^{q\to\gamma}_{\rm non-coll}}{d^3p_0 dt_0}  \,.
\ee

\begin{figure}
  \begin{center}
  \includegraphics[width=0.6\textwidth]{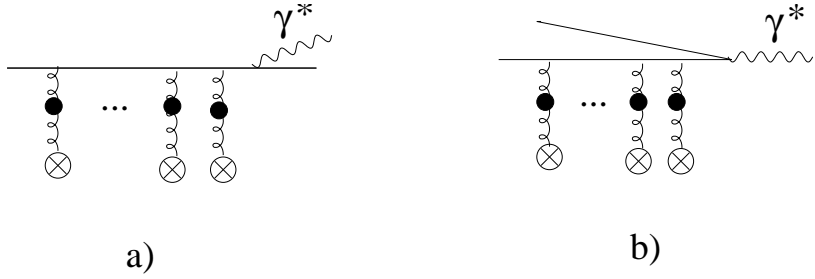}
  \end{center}
  \caption{\label{brem}  Bremsstrahlung and annihilation processes with LPM effect. }
\end{figure}

The retarded photon self-energy $\Pi_\mu^{R\,\mu}$ for collinear processes with LPM effects, i.e. including an infinite sum of diagrams with different number of scatterings with soft gluons (see Fig.~\ref{brem}), has been extended from real photons~\cite{AMY} to virtual photons in Ref.~\cite{AGMZ}, in the limit $M\ll E$.  The dilepton production induced by jets in collinear processes has finally been calculated in Ref.~\cite{Turbide:2007fd} for a longitudinal expanding QGP.   The quark to photon transition rates for collinear processes have been extracted from $\Pi_\mu^{R\,\mu}$ using Eq.~(\ref{photon_TFT}) and relativistic kinetic theory:
\begin{equation}
E_0\frac{dR^{\gamma^*}_{\rm coll}}{d^3p_0}=12 \int_0^\infty dq_{0}\, \frac{q_{0}^2\, E_0}{(2\pi)^3 p_0^2}\, n_{\rm FD}(q_{0}) \left.\frac{d\Gamma^{q\to\gamma}_{\rm coll}(q_{0},p_0,M)}{dp_0  dt_0}\right|_T \left(1-\frac{\theta(p_0-q_{0})}{2}  \right)\,,
\end{equation}
where $n_{\rm FD}$ is the Fermi-Dirac phase-space distribution function.  The $\theta(p_0-q_{0})$ function is included to avoid double counting in the annihilation process, since $n_{\rm FD}^q(q_{0})n_{\rm FD}^{\bar{q}}(p_0-q_{0})=n_{\rm FD}^q(p_0-q_{0})n_{\rm FD}^{\bar{q}}(q_{0})$.  After substracting from $d\Gamma^{q\to\gamma}_{\rm coll}$ the leading order annihilation (Fig.~\ref{direct}a), again to avoid double counting, the photon yield from the collinear processes ($\Omega_{\rm jet}\approx\Omega_\gamma$ and $ y_{\rm jet}\approx y_\gamma=y$ for $M\ll p_T$) induced by jets in an expanding medium is finally
\be
\label{yield_coll}
E\frac{dN^{\gamma^*}_{\rm coll}}{d^3p}(M) &=&\int dt \int d^2r_\perp\,{\cal P}({\bf r}_\perp)\int dq\, \frac{q\,E}{p^2}\, \frac{dN^{q\bar{q}}_{\rm jet}(t,b)}{d^2q_{T}\, dy}\,{\cal J}\frac{d\Gamma^{q\to\gamma}_{\rm coll}(q_{0},p_0,M)}{dp_0  dt_0}\,.
\ee

The Jacobian of the $dt_0\,dp_0\to dt\,dp$ transformation is 
\be
{\cal J}=\frac{E_0}{p_0}\left(\frac{p}{E}-\frac{\pv\cdot\bet}{p} \right)\,.
\ee

\subsection{Thermally induced processes}

The photons produced during the thermal phase due to the collisions of thermal particles, are calculated simply by
\be
E\frac{dN^{\gamma^*}_{\rm thermal}}{d^3p}(M)&=&\int d^4x E_0\frac{dR^{\gamma^*}_{\rm thermal}}{d^3p_0}(M),
\ee
where the photon production rate is evaluated in the local thermal frame.  The QGP induced processes also correspond to the diagrams shown in Figs.~\ref{direct} and ~\ref{brem}, with the difference that the incoming particles are now thermal partons rather that jets. The production rates for the QGP induced photons are from Refs.~\cite{Turbide:2006mc} and ~\cite{AGMZ}, while the hadronic gas (HG) production rates (mesonic as well as baryonic) used here are introduced in Ref.~\cite{TRG}.  
Defining $f_{\rm QGP}$ as the QGP content in the mixed phase, the thermally induced radiation is
\be
E\frac{dN^{\gamma^*}_{\rm thermal}}{d^3p}(M)=\int d\tau\tau d\eta d^2{\bf x}_\perp && \left[f_{\rm QGP}\,E_0\frac{dR^{\gamma^*}_{\rm QGP}}{d^3p_0}(M)\right.\\ \nonumber &&+\left.(1-f_{\rm QGP})E_0\frac{dR^{\gamma^*}_{\rm HG}}{d^3p_0}(M)\right]\,\Theta(T(\tau,\eta,{\bf x}_\perp)-T_{\mbox{dec}})\,,
\ee
where $T_{\mbox{dec}}$ denotes the decoupling temperature.  

\subsection{Prompt photon production}

\begin{figure}
  \begin{center}
  \includegraphics*[width=4in]{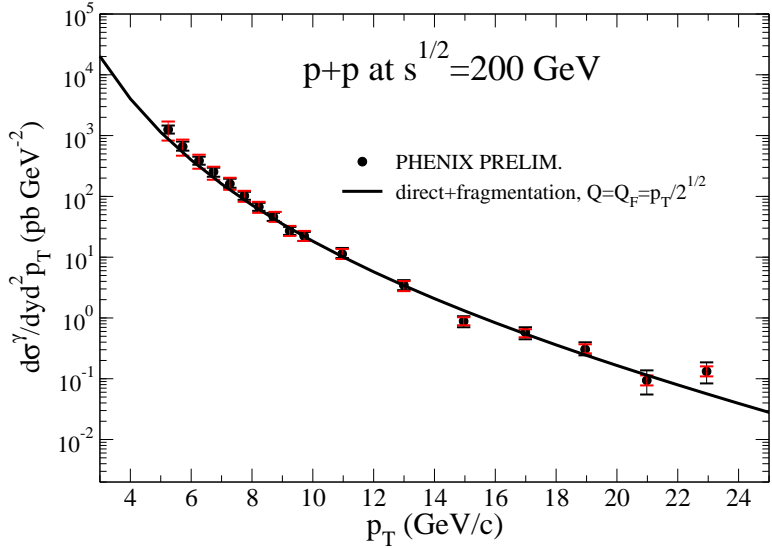}
  \end{center}
  \caption{\label{prompt_pp} (Color online)  Distribution of real prompt photons in $p-p$ collisions at RHIC. Data points are from PHENIX~\cite{Adler:2006yt}.}
\end{figure}

The prompt photon production in $p-p$ collisions is given by
\be
\label{prompt}
E\frac{d\sigma^{pp}_{\rm prompt}}{d^3p}(M)&=&\frac{d\sigma^{pp}_{\rm dir}}{d^2p_T dy}(M,Q,Q_F)+\frac{d\sigma^{pp}_{\rm frag}}{d^2p_T dy}(M,Q,Q_F)\nonumber\\
&=&\sum_{a,b,c}\int dx_a dx_b f_a(x_a,Q)f_b(x_b,Q)K_{\rm dir}(p,M,Q,Q_F) E\frac{d\sigma_{a+b\to c+\gamma^*}(p,M,Q)}{d^3p}\nonumber \\ &+& \sum_{a,b,d}\int dx_a dx_b f_a(x_a,Q)f_b(x_b,Q)\int \frac{dz}{z^2} K_{\rm frag}(p,M,Q,Q_F)\nonumber \\ &&\times\left.\frac{d\sigma_{a+b\to c+d}(M,Q)}{d^2q_T dy}\right|_{q_T=p_T/z} D_{\gamma^*/c}(M,z,Q_F)\,. 
\ee
Here, the renormalization scale has been implicitly set equal to the factorization scale $Q$ and the leading order expression for the cross-section $d\sigma_{a+b\to c+\gamma^*}$ and $d\sigma_{a+b\to c+d}$ can be found in Refs~\cite{guo} and~\cite{owens}.  The $K$-factors include NLO effects. We assume, for $M\ll p_T$, that $K(M,p,Q,Q_F)\approx K(p,Q,Q_F)$, and we evalutate them using the numerical program from Aurenche {\it et al.}~\cite{aurenche}.  The real photon vacuum fragmentation function comes from Ref.~\cite{photonfrag}  while for $M\ne$0, we take the leading order result~\cite{berger}. The splitting between the direct and fragmentation contributions is arbitrary and depends on the choice of fragmentation scale $Q_F$.  So, only the sum of the two contributions has a clear interpretation.  We nevertheless set $Q=Q_F$, and ajust $Q_F$ to fit the recent PHENIX prompt (real) photon results in $p-p$~\cite{Adler:2006yt}.  Using $Q=p_T/\sqrt{2}$, we obtain a nice agreement with data, as shown in Fig.~\ref{prompt_pp}.

In nucleus-nucleus collisions, the direct contribution is simply given by
\be
E\frac{dN^{\gamma^*}_{\rm dir}}{d^3p}(M,Q,Q_F,b)= T_{AB}(b) \left.\frac{d\sigma^{pp}_{\rm dir}}{d^2p_T dy}(M,Q,Q_F)\right|_{f_a\to g_a}\,,
\ee
where the pdf functions $f_a$ are replaced by $g_a$, defined in Eq.~(\ref{iso_sha}),  to include isospin and shadowing effects. The fragmentation contribution, like the jet-fragmentation contribution in pion production, could suffer high-$p_T$ suppression due to the quenching of jets in the QGP.  Remember that the distribution of jets has been calculated in Eq.~(\ref{jet_initial}) by setting $Q=q_T$, while in Eq.~(\ref{prompt}), the scale has been set to $Q=p_T/\sqrt{2}$ .  Thus, in order to make a connection with the distribution of jets, we approximate the fragmentation contribution in nucleus-nucleus collisions as
\be
E\frac{dN^{\gamma^*}_{\rm frag}}{d^3p}(M,Q,Q_F,b)&\approx &\sum_{i=q\bar{q},g} \int d^2r_\perp\,{\cal P}({\bf r}_\perp)\int dq_T\frac{1}{p_T} \frac{dN_{\rm jet}^i({\bf d},Q^{'}=q_T,b)}{d^2q_{T} dy}\nonumber \\&&\times H(q_T,Q^{'},Q)\, D_{\gamma^*/i}(M,z=p_T/q_T,Q_F)\,.
\ee
The effective function $H(q_T,Q^{'},Q)$ has been introduced so that if the jet is not quenched in the medium, we get the following:
\be
\left.E\frac{dN^{\gamma^*}_{\rm frag}}{d^3p}(M,Q,Q_F,b)\right|_{no\,E-loss}=T_{AB}(b)\left.\frac{d\sigma^{pp}_{\rm frag}}{d^2p_T dy}(M,Q,Q_F)\right|_{f_a\to g_a}\,.
\ee
For $Q=Q_F=p_T/\sqrt{2}$ and $Q^{'}=q_T$, we obtain that $H(q_T,Q^{'},Q)\sim 1.9$. The quenching of jets in the medium will depend on the path lenght ${\bf d}$, which depend on the propagating direction of jets and the position ${\bf r}_\perp$ in the transverse plane where they have been created.

\section{Dilepton and real photon production}
\label{sec_pro}

The total direct photon yield is the sum of all contributions discussed in the preceding sections:
\be
\frac{dN^{\gamma^*}_{\rm total}}{d^2p_T dy}(M,b)=E\frac{dN^{\gamma^*}_{\rm non-coll}}{d^3p}+E\frac{dN^{\gamma^*}_{\rm coll}}{d^3p}+E\frac{dN^{\gamma^*}_{\rm thermal}}{d^3p}+E\frac{dN^{\gamma^*}_{\rm dir}}{d^3p}+E\frac{dN^{\gamma^*}_{\rm frag}}{d^3p}\,,
\ee
where $d^2p_T=p_T\,dp_T\,d\phi$. The real photon spectrum is simply obtained by
\be
\frac{dN^{\gamma}_{\rm total}}{d^2p_Tdy}(b)=\frac{dN^{\gamma^*}_{\rm total}}{d^2p_T dy}(M=0,b)\,,
\ee
while the dilepton spectrum is
\be
\frac{dN^{e^+e^-}_{\rm total}}{dM^2d^2p_Tdy}(|y_{e^\pm}|\le y_{\rm cut},b)=\frac{\alpha}{3\pi M^2}\frac{dN^{\gamma^*}_{\rm total}}{d^2p_T dy}(M,b)P(|y_{e^\pm}|\le y_{\rm cut},p_T,M)\,.
\ee
The multiplicative factor $P(|y_{e^\pm}|\le y_{\rm cut})$, defined in Ref.~\cite{Turbide:2006mc}, is introduced in order to take care of geometrical acceptance of any detector.  Finally, on top of the photon yield, the nuclear modification factor and the azimuthal anisotropy coefficient provide important informations about behaviour of jets in the medium.  They are respectively defined by
\be
R^{\gamma}_{AA}(b,p_T,y)=\frac{\int_0^{2\pi}\,d\phi\, dN^\gamma(b)/d^2p_Tdy}{2\pi\,T_{AB}(b)\, d\sigma^{pp}_{\rm prompt}/d^2p_Tdy }
\ee
and 
\be
v^\gamma_2(b,p_T,y)=\frac{\int_0^{2\pi}d\phi\,\mbox{cos}2\phi\, dN^\gamma(b)/d^2p_Tdy}{\int_0^{2\pi}d\phi\,dN^\gamma(b)/d^2p_Tdy}\,.
\label{v2}
\ee
In this study, we examine only the production of mid-rapidity photons, such that $y=0$ is set in all the above equations.

\section{Results}
\label{sec_results}

The thermal fluid dynamical evolution is described by  by the longitudinally
boost-invariant (2+1)-dimensional hydrodynamic code AZHYDRO
~\cite{KH,AZHYDRO} with the EOS~Q equation of state which matches
a non-interacting QGP above $T_c$ to a chemically equilibrated hadron 
resonance gas below $T_c$ at the critical temperature $T_c{\,=\,}164$\,MeV. 
We assume an early start of the hydrodynamic evolution at 
$\tau_i$=0.2\,fm/$c$ in order to be able to account schematically 
for the pre-thermalized stage (for
initial conditions see \cite{v2_thermal_ph}).  The 
decoupling temperature is set to $T_\mathrm{dec}\approx 
130$\,MeV. In the longitudinal invariance scenario, the temperature of the medium at any space-time point is defined by the proper time and the radial position, such that $T=T(\tau,{\bf x}_\perp)$.  The flow velocity for any space-time rapidity value can be extracted from the flow at $\eta=0$.  Indeed, for $\bet=(\bet_\perp,\beta_z)$, we have
\be
\bet_\perp(\tau,\eta,{\bf x}_\perp)=\frac{\bet_\perp(\tau,\eta=0,{\bf x}_\perp)}{\mbox{cosh}\eta}\,,\quad \beta_z(\tau,\eta,{\bf x}_\perp)=\mbox{tanh}\eta\,.
\ee

The calculated photon spectra, and their different components, are shown in Fig. \ref{photon_yield}. The data are for Au-Au collisions, at the top RHIC energy, for two different centrality classes. The $0 - 10 \%$ and $0 - 20 \%$ classes are shown in the left and right panel, respectively. Note the data in the larger class extends to lower $p_{T}$ than that for the more central class, owing to a different experimental extraction technique \cite{qm2005}. Considering first that figure (left panel), the different contributions highlighted are those from hard primordial scatterings (prompt), which include the photons from Compton and annihilation events, together wit those from the fragmentation of jets. The photon spectrum associated with the interaction of jets with the thermal components of the quark-gluon plasma is labeled jet-QGP.  The radiation from the thermal components of the quark-gluon plasma is shown, together with that from the thermal components of the hot gas of composite hadrons.  The sum of the different contributions is the solid curve; the data are from PHENIX \cite{qm2005}. It is seen that, for the physical case under consideration here, the 
\begin{figure}[!ht]
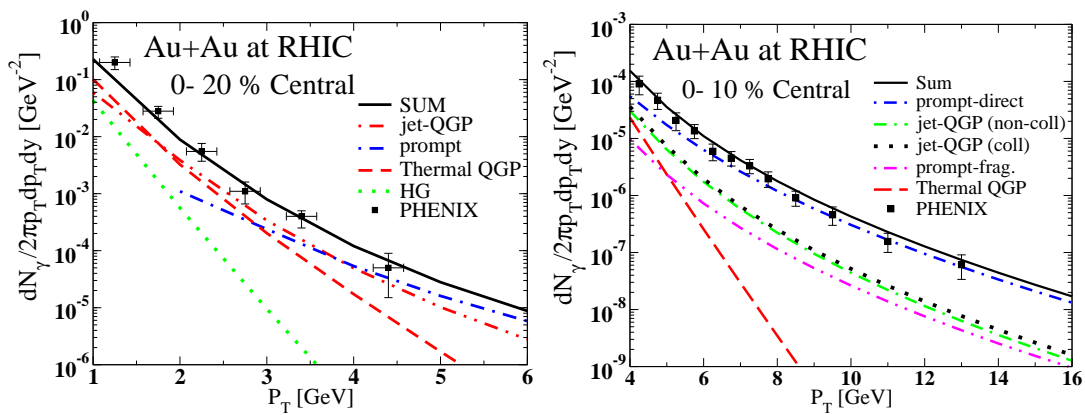

  \begin{center}
  \includegraphics*[width=7cm]{photon_020_2D_new.eps}
 \includegraphics*[width=7.2cm]{photon_010_2D.eps}
  \end{center}
  \caption{\label{photon_yield} (Color online)  Yield of photons in Au+Au collisions at RHIC, for two centrality classes: 0-20$\%$  (left panel) and 0-10$\%$ (right panel). The different elements of the theoretical calculation are described in the text. The data are from Refs. \cite{qm2005}, and \cite{Isobe:2007ku}, respectively. } 
 \label{photon_yield}
\end{figure}
jet-plasma photons are important to the theoretical interpretation of the experimental data in the window $2 < p_T <  4$ GeV. For smaller values of $p_{T}$, the emission from thermal media (whether QGP or hadron gas) represents a sizeable source. For the higher transverse momentum data, the radiation from hard collisions gradually take over the whole spectrum. This picture receives additional support from the higher $p_{T}$ data in the right panel. Most of that data is dominated by Compton and annihilation contributions calculated from pQCD. The jet-plasma sources are demanded only by the first two data points. The purely thermal contributions are subdominant in the entire range spanned by this figure. The fragmentation contribution to the real photon spectrum is small, owing mainly to the energy lost by the propagating jets. Here again, adding all of the  sources produces  a signal in agreement with the measured data.

\begin{figure}[h]
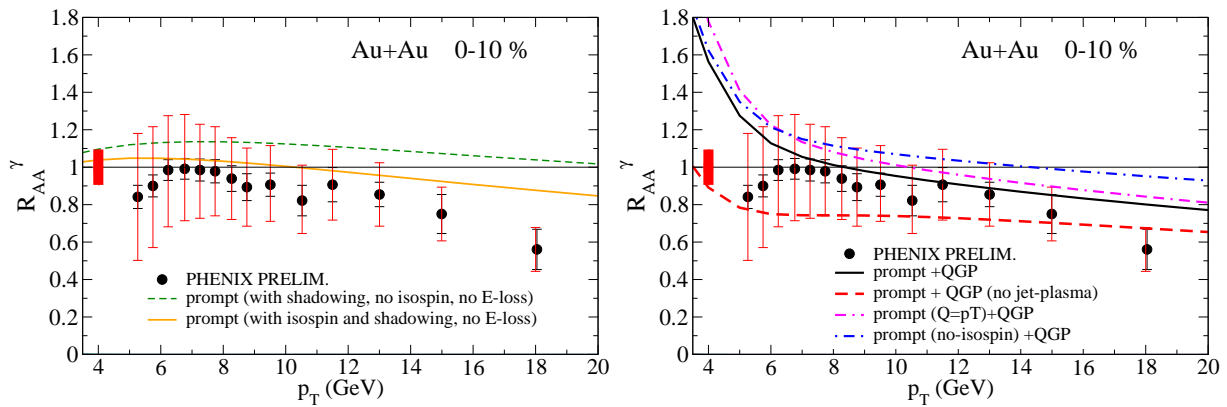

  \begin{center}
  \includegraphics*[width=8cm]{R_AA_phot_1.eps}
  \includegraphics*[width=8cm]{Raa_new_photon.eps}
  \end{center}
  \caption{\label{Raa_photon} (Color online)  Nuclear modification factor of direct photon in central Au+Au collisions at RHIC in 2D+1 hydro, with a scale $Q=p_T/\sqrt{2}$ in the prompt contribution.  Left panel: effect of shadowing and isospin on the prompt contribution without medium effects.  Righ panel: the effect of QGP and the scale is studied.  The effect of a scale $Q=p_T$ is shown by the double dash-dotted line, while the effect of removing all photons produced from jet-medium interactions is shown by the dashed line.  The result obtained without isospin effects is shown by the dot-dashed line.   Data points are from PHENIX~\cite{Isobe:2007ku}.}
\label{photonRAA}
\end{figure}

Another useful representation of the experimental data and a quantitative measure of the nuclear effects is provided by a plot of $R_{AA}^{\gamma}$ (for real photons), shown in Fig. \ref{photonRAA}. The experimental data seem to show an interesting trend pointing towards diminishing values of $R_{AA}^{\gamma}$ as $p_{T}$ grows. The experimental error bars are too large to permit a precise quantitative assessment, but different interesting possibilities and combinations may be considered. 
The left panel of Fig. \ref{photonRAA} shows $R_{AA}^{\gamma}$ calculated under different assumptions, and basically shows the importance of cold nuclear effects. The dashed curve shows the effect of the nuclear environment (shadowing) on the parton distribution function (pdf), while neglecting the specific isospin composition of the colliding nuclei. The full curve includes both isospin and shadowing contributions. The results of both calculations are systematically higher than the experimental data centroids, and exhibit a smaller slope than the one seen in the measurements, although the isospin effect can cause a 20$\%$ reduction at high-$p_T$, as also found  in Ref.~\cite{Arleo:2006xb}. The right panel includes medium effects, calculated as described earlier in the text; all curves except one contain jet-plasma photons, together with leading parton energy loss as evaluated with AMY. The dashed-dotted line shows the effect of neglecting the isospin content of the parton distribution functions. The double-dash dotted curve shows the scale-dependence of $R_{AA}^{\gamma}$, with the result of using $Q=p_{T}$ for the prompt contribution instead of $Q=p_{T}/\sqrt{2}$ used elsewhere in this work. The full curve shows the nuclear modification factor evaluated with all sources described in this paper, together with the relativistic hydrodynamics evolution. Recall that the relativistic hydrodynamics modeling is constrained by a set of soft hadronic data \cite{KH}. The larger visible effect on the nuclear modification factor appears when jet-plasma photons are neglected (dashed line), causing a~$30\%$ reduction at $p_T=8$ GeV.  The jets are however allowed to loose energy before fragmentation (like all cases in this panel).  Because of the large errors, the data does not currently permit to choose between the cases where the jet-plasma photons are present or absent. However, it is important to realize that $R_{AA}^{\gamma}  <  1$ at higher values of $p_{T}$, is a direct consequences of the fragmentation photons being affected by the energy loss of the fragmenting jet, as well as isospin effect in the nucleus-pdf. Should this trend, apparent in Figure \ref{photonRAA}, be confirmed experimentally, a quantitative link would exist between the high momentum nuclear modification factor of photons, and that of strongly interacting particles also born out of jet fragmentation. It is important for the same approach to reproduce both observables. Also, the large values of $R_{AA}^{\gamma}$ observed at $p_{T} < 6 $GeV/c (right panel of Fig. \ref{Raa_photon}) are  directly attributable to thermally-induced channels, in our approach. Our calculated results appear to overestimate the central values of the measured quantities (note however that the denominator of $R_{AA}^{\gamma}$ is slightly underestimated at low $p_{T}$ by pQCD: correcting this will make our result correspondingly smaller), but smaller error bars would go a long in quantifying the medium-related processes.

\begin{figure}[h]
  \begin{center}
  \includegraphics*[width=8cm]{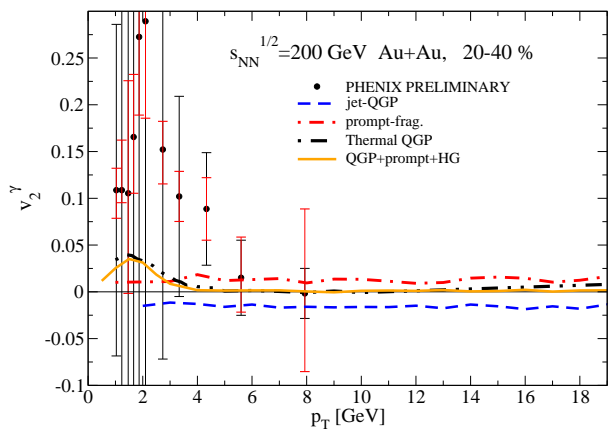}
  \end{center}
  \caption{\label{v2_fig} (Color online) Azimuthal anisotropy of direct photons in 20-40$\%$ central collisions at RHIC, within a 2D+1 hydro model. Dashed line : jet-plasma contributions; dot-dashed line: jet-fragmentation contribution; double dot-dashed line: thermal radiation of QGP; solid line: sum of QGP, prompt and hadronic gas contributions. The data are from Ref. \cite{Sakaguchi:2007zs}.}
\end{figure}
We turn now to calculations and measurements of photon azimuthal anisotropy. This was discussed for low $p_T$ photons in Ref. \cite{v2_thermal_ph}, and for high $p_{T}$ photons in Ref. \cite{TGF2006}; both regions are treated here. Using Eq.~(\ref{v2}), $v^{\gamma}_{2}$ (for real photons) can be calculated for the different ingredients of the theoretical treatment, and compared to experimental data. This is done in Fig. \ref{v2_fig}.  One observes that the net anisotropy (full line) is very low and in fact essentially zero for $p_{T} >$ 4 GeV. For smaller values of the transverse momentum, our results are smaller than the central values of the experimental data (even if the error bars are large). The $v_{2}$ coefficient for the photons originating from the jet-plasma interactions is indeed negative \cite{TGF2006}, but its magnitude is numerically smaller than in these earlier estimates. The explanation for this difference is two-fold. First, the present calculations relies on a realistic 2D+1 modeling \cite{KH} of the space-time evolution of the hot and dense medium, as opposed to using a simpler 1D Bjorken expansion. 
\begin{figure}[h]
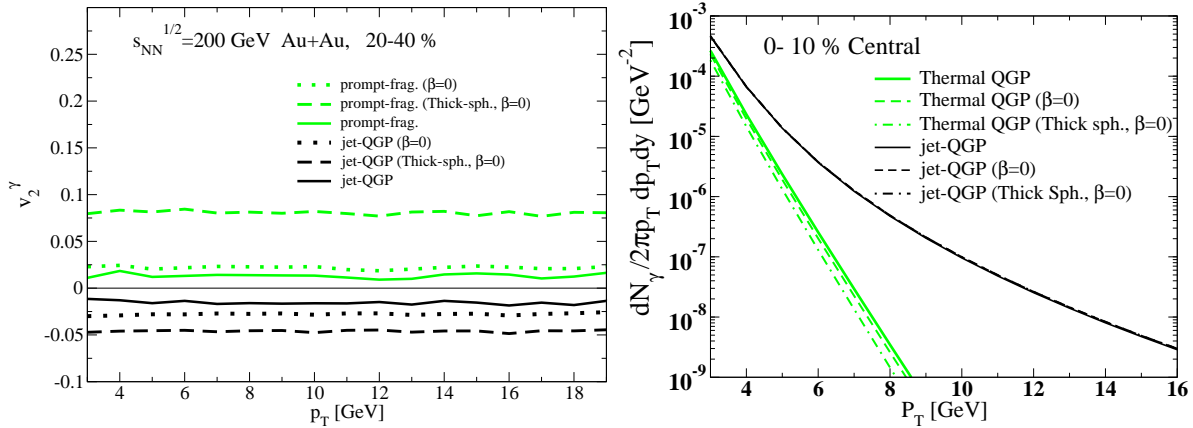

  \begin{center}
\includegraphics*[width=8cm]{v2_geom_effects.eps}
\includegraphics*[width=7.6cm]{photon_2d_vs_1d.eps}
  \end{center}
  \caption{\label{photon_beta_effect} (Color online) The effect of flow and of the geometry of the  initial jet-source profile, on the azimuthal anisotropy (left panel) and on the spectrum of photons (right panel).}
  \label{geom}
\end{figure}
The second reason has to do with the extreme sensitivity of $v^\gamma_{2}$ on the initial conditions of the cooling and expanding source. This is seen in Figure \ref{geom}. The geometry of the initial jet profile ($T_{AB}$) is calculated here using a realistic Woods-Saxon distribution, unless noted otherwise. In the alternate case, thick spheres were used to calculate the nuclear overlap. A general observation is that $v^\gamma_{2}$ is essentially flat as a function of $p_{T}$, for all cases studied here. Considering first the jet-fragmentation contribution (in other words, the fragmentation contribution of the prompt production), the azimuthal anisotropy is positive and larger for the thick sphere geometry, and no radial flow. In the no-flow exercise, $\bet$ is set to 0 and the temperature at each point of the medium evolves from the initial profile $T_i({\bf r}_\perp)$ according to a Bjorken one-dimensional expansion $T({\bf r}_\perp)=T_i({\bf r}_\perp)(\tau_i/\tau)^{1/3}$. The Woods-Saxon distribution, together with the flow bring down $v^\gamma_{2}$ to a level of roughly 2\%. The situation is similar for the jet-plasma component, the azimuthal anisotropy has a similar magnitude, but with an overall negative sign. The consequences of ignoring the radial flow and of varying the geometrical profile are negligible for the spectrum of jet-plasma photons, as seen from the right panel of Figure \ref{geom}.  While the high-$p_T$ QGP thermal photons, produced early in the medium evolution, are not much affected by the presence of transverse flow, they are largely affected (by a factor 3) by the choice of the geometrical profile which determine the temperature profile in the transverse plane, to which the thermal production rates are sensitive. At low-$p_T$, thermal radiations dominate the spectrum and bring the total photon anisotropy coefficient up to 3-4$\%$ (Fig.\ref{v2_fig}).  This behaviour is however difficult to validate with the experimental data, owing again to the large size of the error bars.

In summary, the azimuthal anisotropy is smaller (closer to zero) using a 2D+1 hydrodynamic model with realistic geometry than using a 1D Bjorken expansion with a thick sphere initial jet distribution. The flow dynamics and the Woods-Saxon profile conspire to create a smaller geometrical anisotropy as seen by the traveling jets. However, these results do show that the photon $v_{2}$ is rather sensitive to early time dynamics in relativistic nuclear collisions: precise measurements have the potential to stringently constrain evolution approaches.  
\begin{figure}[h]
  \begin{center}
  \includegraphics*[width=8cm]{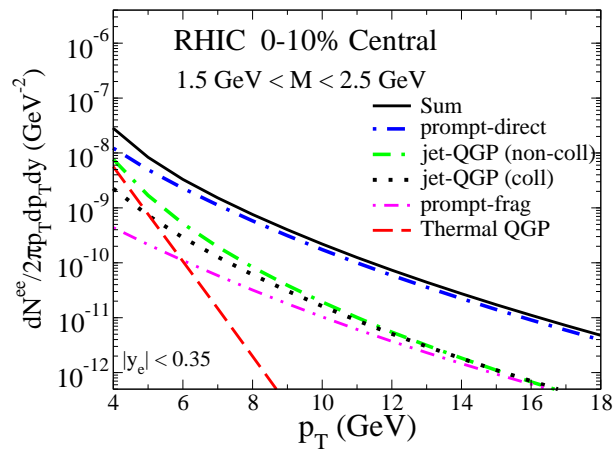}
  \end{center}
  \caption{\label{dilep_1525} (Color online) Yield of dileptons in central 0-10$\%$ collisions at RHIC, within a 2D+1 hydro model. See main text for details. }
  \label{dilep}
\end{figure}

Lastly, similar calculations as the ones performed here for photons may be done for high momentum lepton pairs, using the techniques of Ref. \cite{TGSF}. Even though no experimental measurements are yet available for the set of kinematical cuts we apply, it is nevertheless instructive to examine the relative importance and behavior of the different contributions shown in Figure \ref{dilep}. The components involving a plasma contribution are numerically important in the region $ p_{T}  {\rm (GeV)} <$ 8. These include the radiation from the thermal QCD plasma, the effect of jet plasma interactions (with and without the collinear enhancement germane to the many-body treatment). At higher transverse momenta, the spectrum shown here is taken over by the combination of prompt and fragmentation dileptons. We do not show here the pairs coming from the correlated semi-leptonic decay of heavy-quark mesons \cite{TGSF}.


\section{Summary and conclusions}
\label{sec_conclu}

Assuming that the early phases of the relativistic collisions of heavy nuclei produce a quark-gluon plasma which later hadronizes into a hot gas of strongly interacting particles, we have calculated the electromagnetic signature of the different phases of the hot and dense dense, and evaluated the integrated signal. Within present error bars, we find agreement with the preliminary measurements of the PHENIX collaboration for the photon spectra, the photon nuclear modification factor, and the photon azimuthal anisotropy. Importantly, the dynamical evolution of the strongly interacting system is governed by relativistic hydrodynamics with no parameters in addition to the ones needed to quantitatively reproduce a large set of soft hadronic observables. Induced contributions from the QGP influence significantly the spectra, $R_{AA}^{\gamma}$, and elliptic flow of photons for $p_{T} <$ 5 GeV. Our calculations predict larger values for $R_{AA}^{\gamma}$ and smaller values for $v_{2}^{\gamma}$ than suggested by the central values of the experimental data presently available \cite{qm2005}. The latter are, however still subject to considerable experimental uncertainties. At higher $p_{T}$, our results confirm that electromagnetic observables are precise probes of early time dynamics, especially the signal's azimuthal anisotropy in momentum space. According to our calculations, the jet-plasma contributions dominate the photons yield in the window 2$<p_T<$4 GeV.  More precise value about the nuclear modification factor of photons would have the potentiel to discriminate between an inclusive and a non-inclusive jet-plasma contributions scenario. The extension of our findings to LHC energies is under way.

\begin{acknowledgments} 
We thank Fran\c cois Arleo, Sangyong Jeon, Fran\c cois G\'elis,  Guy D. Moore, Guang-You Qin, J\"org Ruppert, and Dinesh K. Srivastava for useful discussions. This work was supported in part by the Natural Sciences and Engineering Research Council of Canada, and in part by the U.S. Department of Energy under Contract No. DE-FG02-01ER41190.
\end{acknowledgments}

\end{document}